\documentclass{aastex}          
\usepackage{spr-astr-addons}    

\begin{document}
%
\title{Stellar abundances of beryllium and CUBES}

\shorttitle{Stellar abundances of beryllium and CUBES}
\shortauthors{R. Smiljanic}

\author{R. Smiljanic\altaffilmark{1}} 

\altaffiltext{1}{Department for Astrophysics, Nicolaus Copernicus Astronomical Center, ul. Rabia\'nska 8, 87-100 Toru\'n, Poland}

\begin{abstract}
Stellar abundances of beryllium are useful in different areas of astrophysics, including studies of the Galactic chemical evolution, of stellar evolution, and of the formation of globular clusters. Determining Be abundances in stars is, however, a challenging endeavor. The two Be II resonance lines useful for abundance analyses are in the near UV, a region strongly affected by atmospheric extinction. CUBES is a new spectrograph planned for the VLT that will be more sensitive than current instruments in the near UV spectral region. It will allow the observation of fainter stars, expanding the number of targets where Be abundances can be determined. Here, a brief review of stellar abundances of Be is presented together with a discussion of science cases for CUBES. In particular, preliminary simulations of CUBES spectra are presented, highlighting its possible impact in investigations of Be abundances of extremely metal-poor stars and of stars in globular clusters.
\end{abstract}


%
\section{Introduction}\label{sec:intro}

Beryllium has one single stable isotope, $^{9}$Be. Spectral lines of Be have been identified in the Sun, in the near UV region that is observable from the ground, quite some time ago by \citet{1895ApJ.....1...14R}. The lines tentatively identified included the Be I lines at $\lambda$ 3321.011, 3321.079 and 3321.340 \AA, and the Be II resonance lines at $\lambda$ 3130.422 and 3131.067 \AA\ \citep[wavelengths from][]{1997JPCRD..26.1185K,2005PhyS...72..309K}. 

However, it has been shown that the identification of the Be I lines was erroneous \citep{1968SoPh....5..159G,1968ApL.....2..235H}. Only the Be II $^{2}$S--$^{2}$P$_{0}$ resonance lines at $\sim$ 3130 \AA\ are now used to determine Be abundances in late-type stars. Other Be lines can be found in the UV below 3000 \AA, but are only observable from space and seem to be of limited use \citep{1996A&A...313..909G}. Isotopic shifted lines of unstable $^{7}$Be and $^{10}$Be have been searched for, but were never detected \citep{1968PASP...80..622W,1973ApJ...185L..27B,1975A&A....42...37C}.

\begin{figure}[t]
\begin{center}
 \includegraphics[width=7cm]{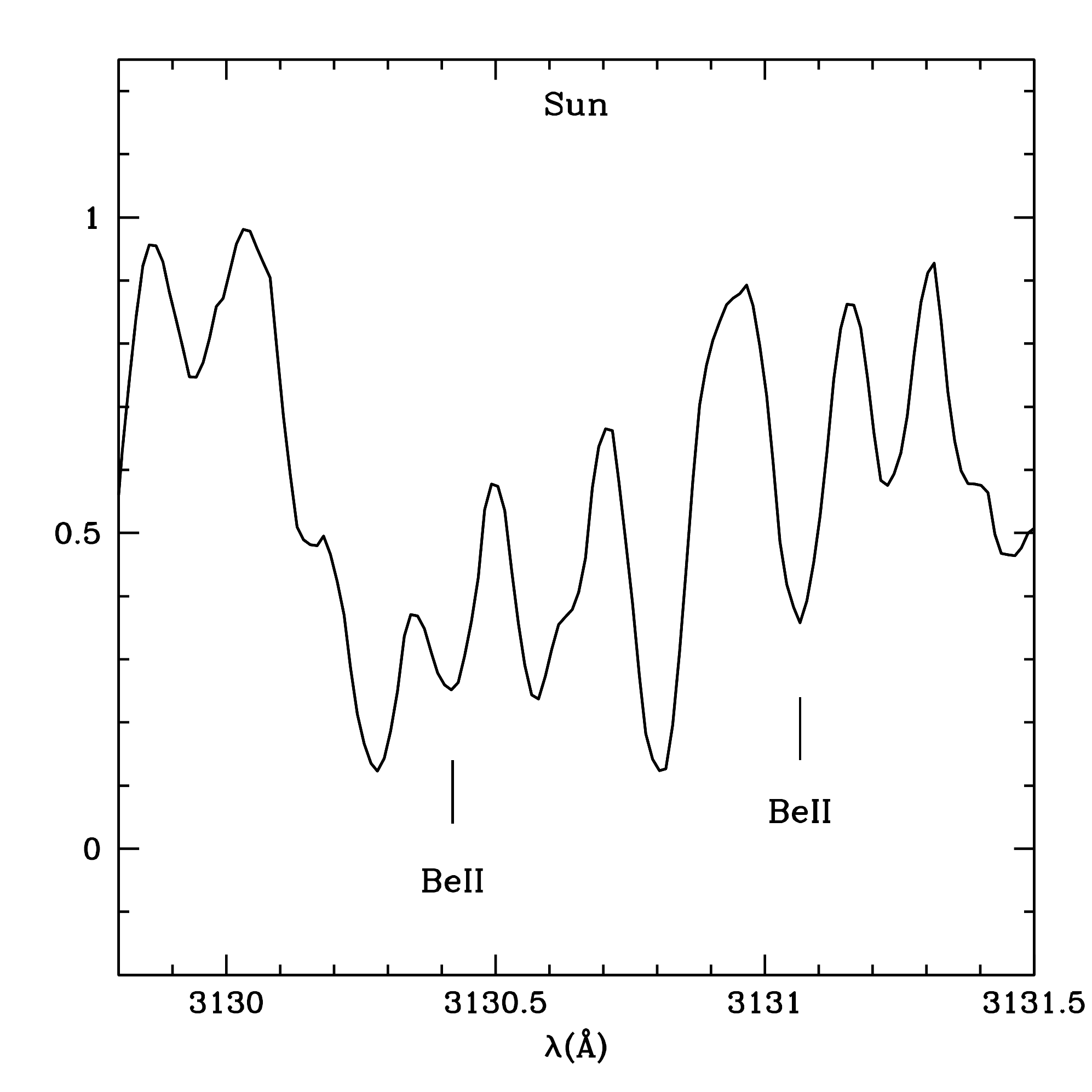}
 \end{center}
 \caption{The Be lines in the solar spectrum} 
 \label{fig:sun}
 \end{figure}

High signal-to-noise (S/N) spectra in the near UV is hard to obtain because of atmospheric extinction. In addition, the spectral region of the Be lines is crowded with other atomic and molecular lines (see the solar spectrum in Fig. \ref{fig:sun}). Some of these blending lines are still unidentified, as for example the one contaminating the blue wing of the 3131.067 \AA\ line \citep[see e.g.][]{1974SoPh...36...11R,1995A&A...302..184G,1997ApJ...480..784P,2011A&A...535A..75S}. 

Early analyses of Be abundances were reviewed by \citet{1969ARA&A...7...99W} and \citet{1976PASP...88..353B}. The solar Be abundance seems to have been first determined by \citet{1929ApJ....70...11R}. The approximation of local thermodynamic equilibrium (LTE) has been shown to result in correct abundances for the Sun \citep{1975A&A....42...37C,1979AJ.....84.1756S,2005ARA&A..43..481A}. The solar Be abundance also seems to be largely insensitive to 3D hydrodynamical effects \citep{2004A&A...417..769A,2009ARA&A..47..481A}. The current value of the solar meteoritic abundance of Be is A(Be)\footnote{The abundance of an element X in this notation is A(X) = $\log \epsilon$(X) = $\log$ [N(X)/N(H)] + 12, i.e. an abundance by number in a scale where the number of hydrogen atoms is 10$^{12}$.} = 1.32 \citep{MakishimaNakamura06,2010ppc..conf..379L} while the photospheric abundance is A(Be) = 1.38 \citep{2009ARA&A..47..481A}. Nevertheless, most 1D LTE model atmosphere analyses of Be in the Sun tend to find values between A(Be) = 1.10 and 1.15. The difference is ascribed to near-UV continuum opacity missing in the computations \citep{1998Natur.392..791B,2001ApJ...546L..65B}.

Beryllium is not produced in significant amounts by the primordial nucleosynthesis, because there are no stable elements with mass number 5 or 8 to act as an intermediate step in synthesising $^{9}$Be \citep{1967ApJ...148....3W,1993ApJ...406..569T,2012ApJ...744..158C}. In addition, Be is rapidly destroyed by proton capture reactions when in regions inside a star with a temperature above $\sim$ 3.5 $\times 10^{6}$ K   \citep{1954ApJ...119..113G,1955PhRv...97.1237S}. Therefore, stars do not produce Be.

In stars like the Sun, for example, Be is only present in the external regions of lower temperature \citep[see Fig.\ 1 of][]{1999ApJ...511..466B}. In evolved stars, where the convective envelope has increased in size and mixed the surface material with the interior, Be, unlike Li, is usually not detected \citep{1977ApJ...214..124B,2006A&A...447..299G,2010A&A...510A..50S}. Beryllium has never been detected in Li-rich giants \citep{1997A&A...321L..37D,1999A&A...345..249C,2005A&A...439..227M}.

Long ago, it was understood that Be can only be produced by the spallation of heavier nuclei, mostly from carbon, nitrogen, and oxygen \citep{1955ApJS....2..167F,1967AnPhy..44..426B}. The only known way to produce significant amounts of Be is by cosmic-ray induced spallation in the interstellar medium (ISM), as first shown by \citet{1970Natur.226..727R} and \citet{1971A&A....15..337M}.

Two channels of cosmic-ray spallation might work to produce Be. In the so-called \emph{direct process}, Be is produced by accelerated protons and $\alpha$-particles that collide with CNO nuclei of the ISM \citep{1975A&A....40...99M,1990ApJ...364..568V,1993ApJ...403..630P}. In the \emph{inverse process}, accelerated CNO nuclei collide with protons and $\alpha$-particles of the ISM \citep{1997ApJ...488..730R,1998ApJ...499..735L,1998A&A...337..714V,2002ApJ...566..252V,2012A&A...542A..67P}.

If the first channel dominates in the early Galaxy, Be should behave as a secondary element, as its production rate would be proportional to the metallicity of the ISM. If the second channel dominates, Be should instead behave as a primary element. The two behaviors can be distinguished by the analysis of Be abundances in metal-poor stars. For primary elements, one should observe a linear correlation between $\log$(Be/H) and the metallicity [Fe/H]\footnote{[A/B] = log [N(A)/N(B)]$_{\rm \star}$ $-$ log
 [N(A)/N(B)]$_{\rm\odot}$} with a slope close to one. For secondary elements, the slope should be around two. 

\begin{figure}[t]
\begin{center}
 \includegraphics[width=7cm]{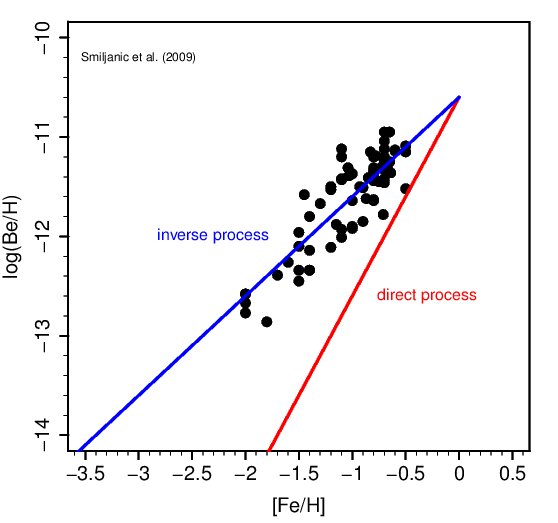}
 \end{center}
 \caption{The Be abundances as a function of metallicty in the metal-poor stars analyzed by \citet{2009A&A...499..103S}. Also shown are the slopes expected for primary and secondary elements. Clearly, the inverse process dominates the Be production and it behaves as a primary element} 
 \label{fig:slopes}
 \end{figure}

Linear relations with the two slopes are compared in Fig. \ref{fig:slopes}. Also shown are the Be abundances of metal-poor stars determined by \citet{2009A&A...499..103S}. It is clear then that Be behaves as a primary element, meaning that the inverse process dominates \citep[as first suggested by][]{1992ApJ...401..584D}. The determination of abundances of Be in metal-poor stars were first attempted by \citet{1984A&A...139..394M} and \citet{1988A&A...193..193R}. But it was with the works of \citet{1992ApJ...388..184R} and \citet{1992Natur.357..379G} that the linear correlation with slope of one became well established. Beryllium abundances in metal-poor stars have been further investigated in several works since then \citep[e.g.][]{1993AJ....106.2309B,1997A&A...319..593M,
1999AJ....117.1549B,2009A&A...499..103S,2009MNRAS.392..205T,2009ApJ...701.1519R,2011ApJ...738L..33T,2011ApJ...743..140B}.

With the inverse process dominating, and considering that cosmic-rays are globally transported across the Galaxy, then the Be production in the early Galaxy should be a widespread process. It follows that, at a given time, the abundance of Be across the Galaxy should have a smaller scatter than the products of stellar nucleosynthesis (such as Fe and O), as suggested by \citet{2000IAUS..198..425B} and \citet{2001ApJ...549..303S}. In this case, Be abundances could be used as a time scale for the early Galaxy \citep{2005A&A...436L..57P,2009A&A...499..103S}. This is an interesting application that still needs to be further constrained.

\section{CUBES}
 
CUBES (Cassegrain U-Band Brazil-ESO Spectrograph) is a new medium-resolution ground based near UV spectrograph for use at the VLT (Very Large Telescope). CUBES is being constructed by Brazilian institutions together with ESO (see Barbuy et al. this volume, for more details). The properties of CUBES are still not finalized, but it will have a resolution of at least R = 20\,000 and should provide access to the wavelength range between 3000--4000 \AA.

CUBES will be more efficient than UVES \citep{2000SPIE.4008..534D} in the near UV. UVES (Ultraviolet and Visual Echelle Spectrograph) is the current spectrograph at the VLT that is able to obtain high-resolution spectra in this region. There is an expected gain of about three magnitudes at 3200 \AA. This gain in sensitivity will expand the number of targets that can be observed. Among the main science cases for CUBES are the study of abundances of Be, C, N, O, and of heavy neutron-capture elements in metal-poor stars (see also Siqueira-Mello, this volume and Bonifacio, this volume). 

Most of the future new instruments and facilities (such as the E-ELT, European Extremely Large Telescope) are now being optimized for the red and near-infrared spectral regions. Thus, CUBES has the potential of being an unique instrument in terms of sensitivity, spectral range, and resolution. With respect to current ESO instruments, CUBES will outperform both UVES and X-Shooter \citep{2011A&A...536A.105V}. CUBES should also be more efficient than other ground-based near-UV capable spectrographs, such as HIRES (High Resolution Echelle Spectrometer) at the Keck Observatory and HDS (High Dispersion Spectrograph) at the Subaru telescope. 

Regarding space-based telescopes, the HST (Hubble Space Telescope) has two UV spectrographs, the Space Telescope Imaging Spectrograph (STIS) and the Cosmic Origins Spectrograph (COS). Nevertheless, the HST will likely not be operational when CUBES comes online ($\sim$ 2017/2018). The WSO-UV (World Space Observatory - Ultraviolet), a 170 cm space based telescope, that should be launched in 2016 will provide access to the UV region (see Shustov, this volume). The WSO-UV is however optimized for the region between 1150-3200 \AA. It is thus not a competitor but complementary to CUBES.

\begin{figure}[t]
 \includegraphics[width=7cm]{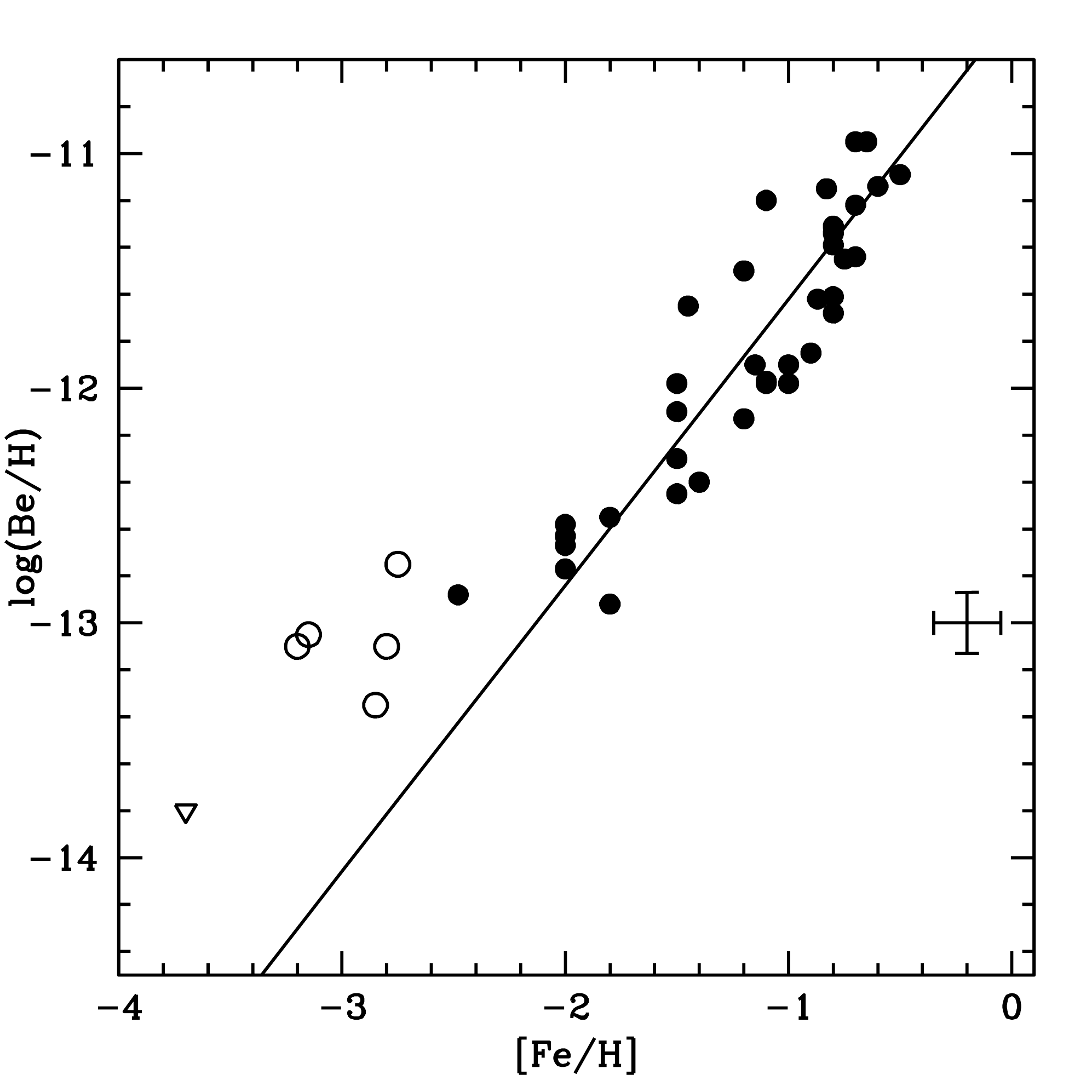}
 \caption{Beryllium abundances as a function of metallicty. The stars from \citet{2009A&A...499..103S} are shown as filled circles, together with stars BD+03 740 ([Fe/H] = $-$2.85), BD$-$13 3442 ([Fe/H] = $-$2.80), G64-12 ([Fe/H] = $-$3.20), G64-37 ([Fe/H] = $-$3.15), and LP 815-43 ([Fe/H] = $-$2.75) as open circles. Also shown is the upper limit for star BD+44 493 ([Fe/H] = $-$3.80), as an open triangle} 
 \label{fig:poor}
 \end{figure}

 \begin{figure*}[t]
\begin{center}
 \includegraphics[width=12cm]{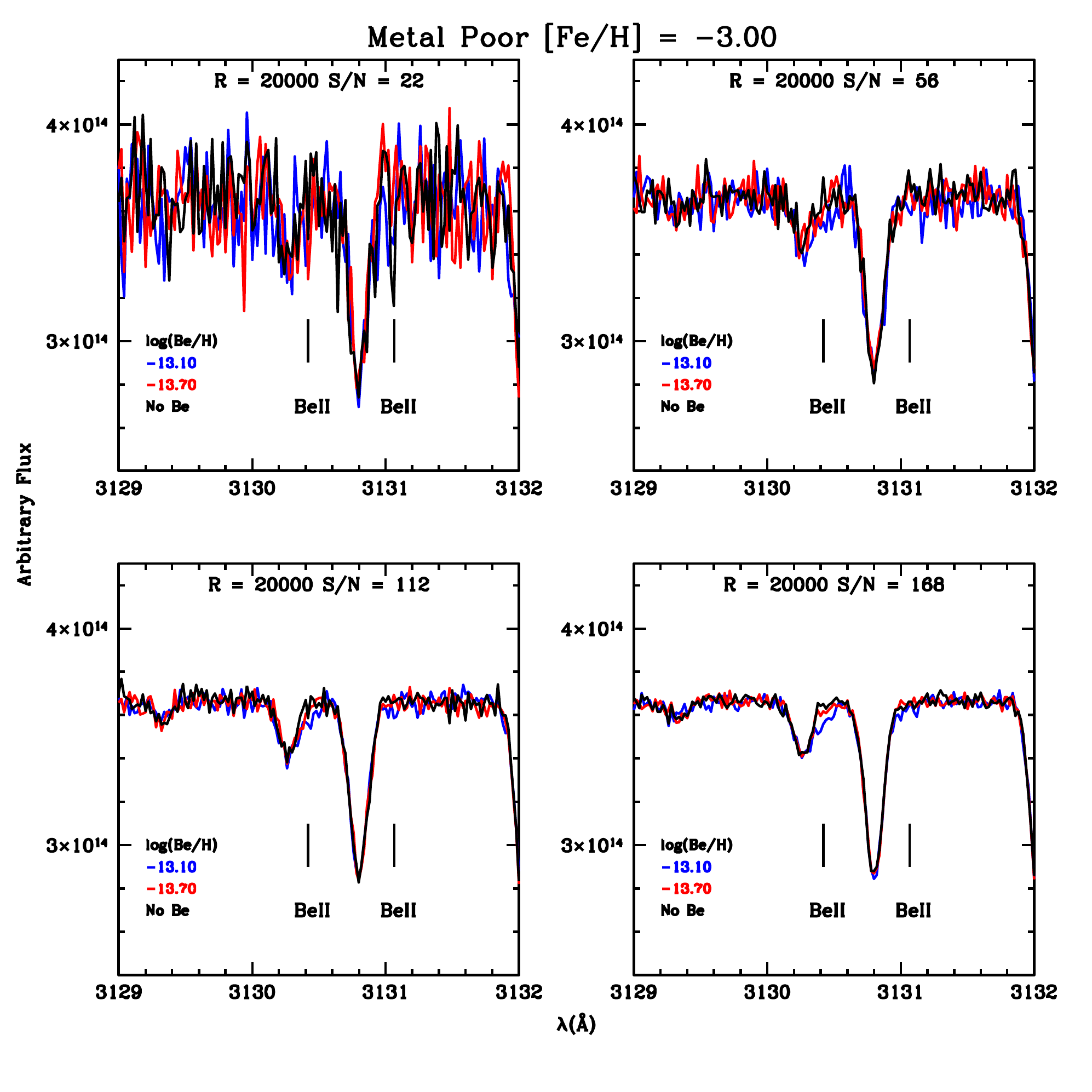}
\end{center}
 \caption{Comparison of three spectra with R = 20\,000 of a metal-poor star with [Fe/H] = $-$3.00. The black spectrum was calculated without Be, the blue with log(Be/H) = $-$13.10 (simulating an abundance plateau), and the red one with an abundance decreased by 0.60 dex. The lines at the red spectra can never be detected} 
 \label{fig:mp1}
 \end{figure*}

\section{Science with beryllium abundances and CUBES}

\subsection{Extremely metal-poor stars}

The linear relation with slope close to one between log(Be/H) and [Fe/H] is well established down to [Fe/H] = $-$2.5/$-$3.0 \citep[but see also][for a slightly different view]{2009ApJ...701.1519R}. Going to the extremely metal-poor regime ([Fe/H] $\leq$ $-$3.00), however, the situation is less clear.

The detection of Be in two stars (LP 815-43 and G64-12) with [Fe/H] $\sim$ $-$3.00 by \citet{2000A&A...362..666P,2000A&A...364L..42P} seems to suggest some deviation from the linear trend,
with a possible flattening of the relation between log(Be/H) and [Fe/H]. The Be abundance of a third star (G64-37), analyzed by
\citet{2006ApJ...641.1122B}, seems to be consistent with this flattening, although these authors argue that there is only evidence
for a dispersion of the Be abundances. The Be abundances of these three stars, together with the stars from \citet{2009A&A...499..103S}, and of stars BD+03 740 and BD$-$13 3442 ([Fe/H] = $-$2.85 and $-$2.80, respectively) are shown in Fig. \ref{fig:poor}. The Be abundances and metallicities of these extremely metal-poor stars were redetermined by \citet{2012ASPC..458...79S}.

A Be upper limit of log(Be/H) $<$ $-$13.80 was determined for the carbon-enhanced metal-poor star BD+44$^{\circ}$ 493 with [Fe/H] = $-$3.80 by \citet{2009ApJ...698L..37I,2013ApJ...773...33I}. This limit is in principle consistent with an extension of the linear trend between Be and metallicity to lower values of Fe (see Fig. \ref{fig:poor}). Care in the interpretation is needed because the origin of the carbon enhancement in this type of stars is also not well established. If any kind of transfer of material from evolved stars (which are depleted in Be) is involved, the surface abundance of Be would be diluted.  

\begin{figure*}[t]
\begin{center}
 \includegraphics[width=12cm]{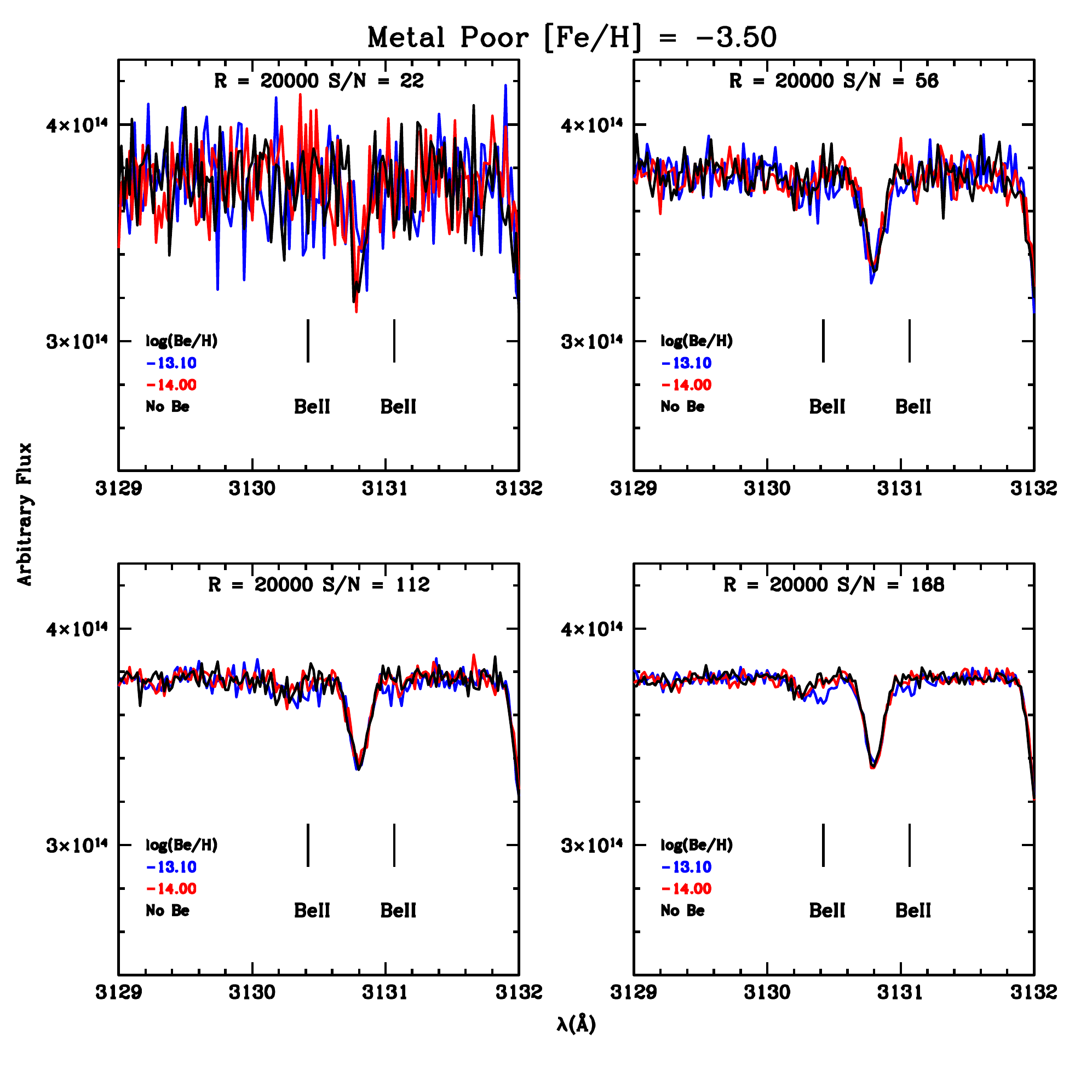}
\end{center}
 \caption{Comparison of three spectra with R = 20\,000 of a metal-poor star with [Fe/H] = $-$3.50. The black spectrum was calculated without Be, the blue with log(Be/H) = $-$13.10 (simulating an abundance plateau), and the red one with an abundance decreased by 1.00 dex. The lines at the red spectra can never be detected} 
 \label{fig:mp2}
 \end{figure*}

Whether the flattening exists or not is thus still not clear, because Be abundances have been determined in only a few stars with metallicity around or below [Fe/H] = $-$3.00 \citep[see][]{2011ApJ...743..140B,2012ASPC..458...79S}. A number of different scenarios could cause such flattening, each one with its own astrophysical implication. \citet{1997ApJ...488..515O}, for example, suggest that a Be plateau, although an order of magnitude below of the current detections, could be the result of an inhomogeneous primordial nucleosynthesis. The identification of such a primordial plateau of Be, similar to the one observed for Li \citep{1982A&A...115..357S}, might be an important test of the standard big bang model. Other possibilities include: 1) the accretion of metal-enriched interstellar gas onto metal-poor halo stars, while crossing the Galactic plane (see \citeauthor{1995ApJ...447..184Y} \citeyear{1995ApJ...447..184Y} for the case of beryllium, but see also \citeauthor{2014ApJ...784..153H} \citeyear{2014ApJ...784..153H} for a more recent assessment of this effect on stellar metallicities); 2) pre-Galactic production by cosmic-rays in the intergalactic medium \citep{2008ApJ...673..676R,2008ApJ...681...18K}; and 3) although a plateau is not predicted, Be enriched stars with [Fe/H] $\geq$ $-$3.2 might be associated to population III quark-novae \citep{2013MNRAS.428..236O}.

To address the existence of the flattening, it is important to expand the number of stars with [Fe/H] $<$ $-$3.00 where Be abundances have been determined. All additional known stars with this or lower metallicity are too faint to be observed with current instrumentation. CUBES will then offer an unique opportunity to obtain the spectra needed to test this possibility.

For this science case, it is important to establish whether a sample of stars has Be abundances at the level of a given plateau or whether the Be abundances follow an extension of  the linear relation between log(Be/H) and [Fe/H]. Therefore, the key requirement is to detect the Be lines at low-metallicities and derive proper abundances, not only upper-limits.

\begin{figure*}[t]
\begin{center}
 \includegraphics[width=12cm]{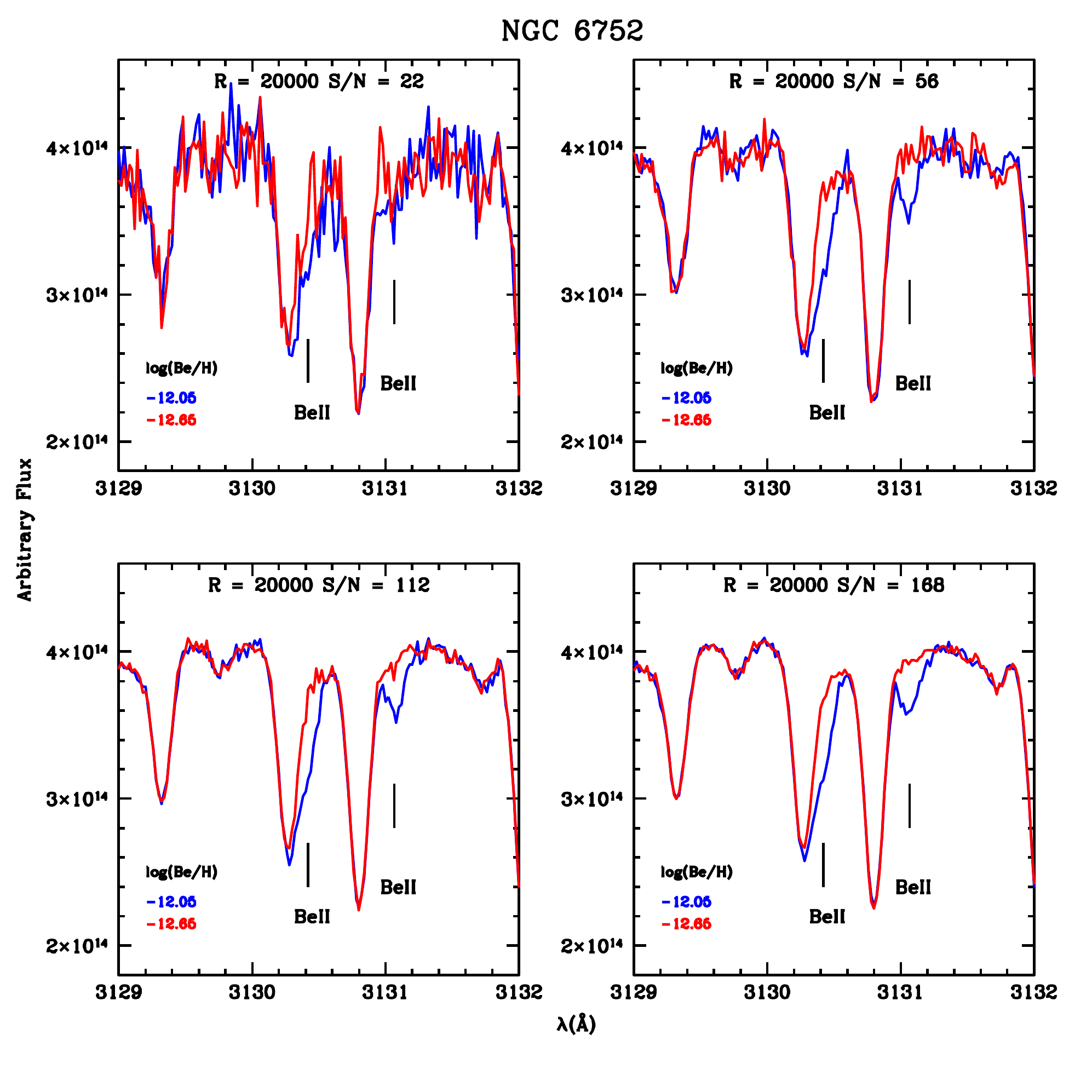}
\end{center}
 \caption{Comparison between turn-off stars of NGC6752 with R = 20\,000. A difference in abundances of $\Delta$Be = 0.60 dex can be detected in all cases except the lowest S/N case, where a detection is marginal at best} 
 \label{fig:n6752}
 \end{figure*}

To test if this will be possible with CUBES, spectra of two different stars were simulated, one with [Fe/H] = $-$3.00 and the other with Fe/H] = $-$3.50. Synthetic spectra were computed with the same codes and line lists used in \citet{2009A&A...499..103S}. The spectra of each star was computed with two different Be abundances. The higher abundance for the two stars is the same, chosen to simulate the presence of a plateau. This abundance is log(Be/H) = $-$13.10, i.e. the Be abundance found for star G64-12 by \citet{2000A&A...364L..42P}. The smaller Be abundance, was decreased by 0.60 dex for the star with [Fe/H] = $-$3.00 and by 1.00 dex for the star with [Fe/H] = $-$3.50:

\begin{enumerate}
\item Metal-poor star 1: $T_{\rm eff}$ = 6300 K; log $g$ = 4.30; [Fe/H] = $-$3.00; $\xi$ = 1.20 km s$^{-1}$
\item Metal-poor star 2: $T_{\rm eff}$ = 6300 K; log $g$ = 4.30; [Fe/H] = $-$3.50; $\xi$ = 1.20 km s$^{-1}$
\end{enumerate}

The Be abundances reflect a scale where A(Be)$_{\odot}$ = 1.10, as derived by \citet{2009A&A...499..103S}. In addition, all the spectra were computed with the abundance of the alpha-elements increased, [$\alpha$/Fe] = +0.40.

The spectra were computed with three different spectral resolutions, R = 15\,000, 20\,000, and 25\,000, although only the case of R = 20\,000 is shown here, as this currently seems to be the favored CUBES value. To simulate the observations, noise was added to the spectra with four different levels in 10 different realizations (kindly computed by H. Kuntschner). The signal-to-noise levels were S/N = 20, 50, 100, and 150 for the case with R = 25\,000. For the other cases the S/N was scaled to try to simulate the gain in S/N given by the decrease in resolution. Thus, the values used were S/N = 22, 56, 112, 168 for the R = 20\,000 case and S/N = 26, 65, 129, 194 for the R = 15\,000 case.

Example are shown in Figs. \ref{fig:mp1} and \ref{fig:mp2}. It is possible to see that detecting low Be abundances (log(Be/H) $\leq$ $-$13.70) is not possible, in the simulated spectra. This is also valid for the cases with different resolutions not shown here. Nevertheless, it is also shown that if there is a plateau around log(Be/H) = $-$13.10, the Be lines can be detected in both metal-poor stars, if S/N $\gtrsim$ 100. Therefore, with CUBES data it would be possible to detect the flattening, if it has a level of Be abundances similar to the one currently found for stars with [Fe/H] $\sim$ $-$3.00. The exact limit of Be detection for CUBES remains to be determined. To calculate that, more information is needed also to understand which level of S/N can be reached.

\begin{figure*}[t]
\begin{center}
 \includegraphics[width=12cm]{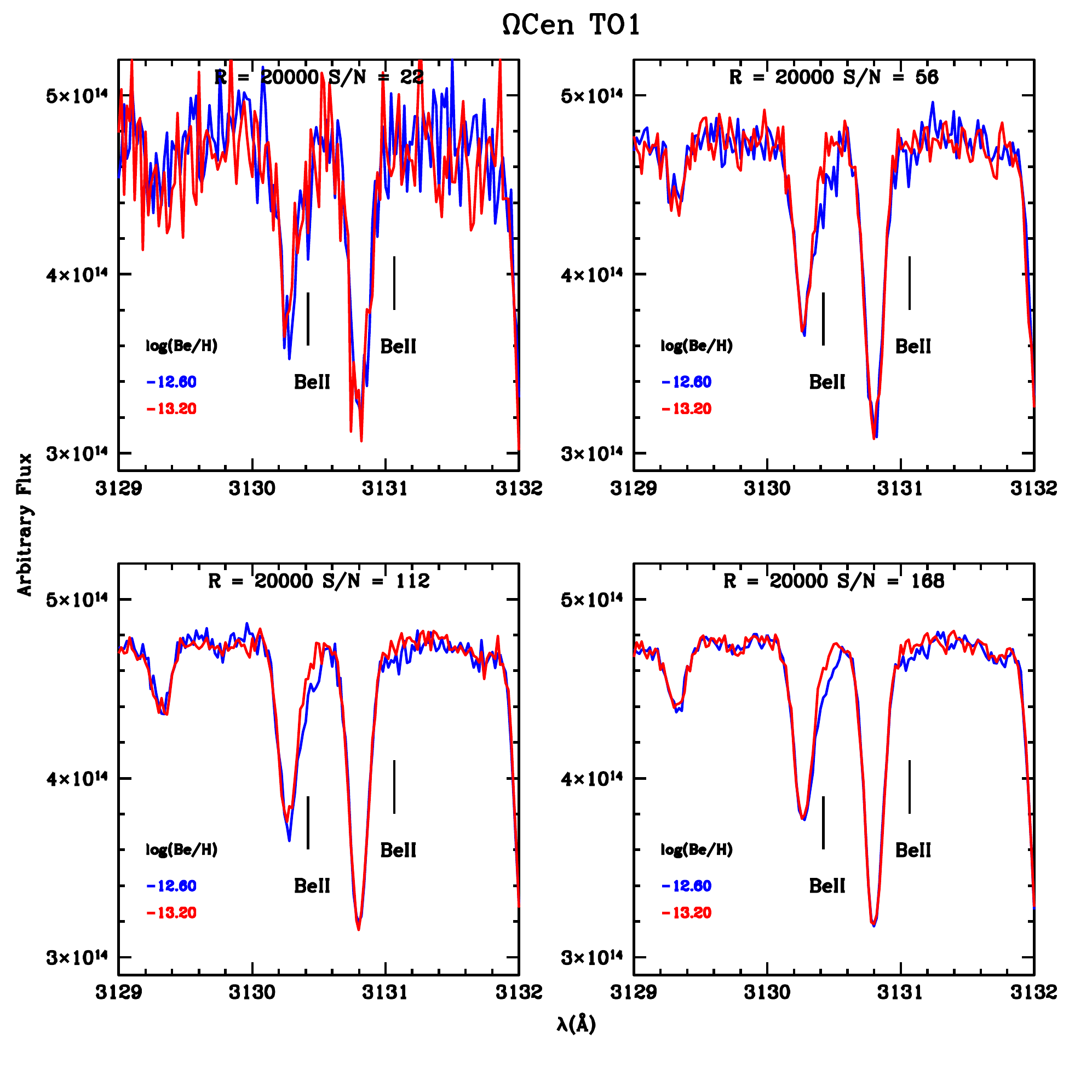}
\end{center}
 \caption{Comparison between stars at the metal-poor turn-off of $\Omega$ Cen with R = 20\,000. A difference in abundances of $\Delta$Be = 0.60 dex can not be easily detected in the lowest S/N case. It seems possible in the three other cases} 
 \label{fig:cen1}
 \end{figure*}

\subsection{Stars in globular clusters}

Globular clusters host multiple generations of stars \citep[see e.g.][and references therein]{2012A&ARv..20...50G}. Second generation stars are formed out of material contaminated by proton-capture nucleosynthetic products. The polluters are usually thought to be either first-generation massive asymptotic-giant-branch stars (AGBs) or first-generation fast-rotating massive stars \citep{2009A&A...499..835V,2007A&A...464.1029D}. 

\begin{figure*}[t]
\begin{center}
 \includegraphics[width=12cm]{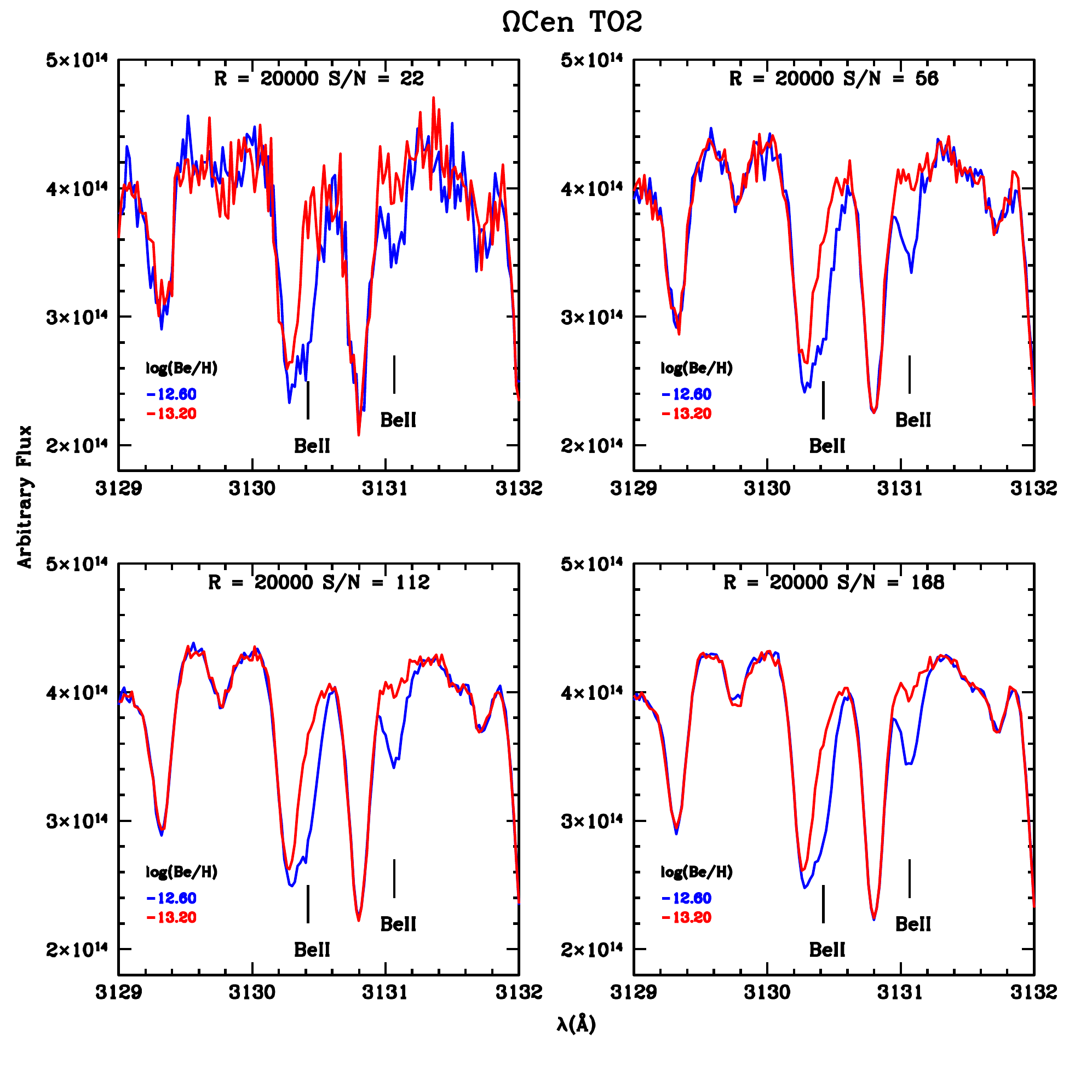}
\end{center}
 \caption{Comparison between stars at the metal-rich turn-off of $\Omega$ Cen with R = 20\,000. A difference in abundances of $\Delta$Be = 0.60 dex can be detected in all cases} 
 \label{fig:cen2}
 \end{figure*}

That these peculiarities have a pristine origin, and are not the result of deep mixing inside the stars themselves, became clear when the same chemical pattern was found in turn-off stars \citep{2001A&A...369...87G}. The most obvious observational result of mixing pristine and processed material is the appearance of the correlations and anti-correlations between light-element abundances. Proton-capture reactions can decrease the abundances of Li, C, O, and Mg and enhance the abundances of N, Na, and Al. All the globular clusters analyzed so far with sufficient depth show signs of these chemical inhomogeneities \citep{2009A&A...505..139C}.

Beryllium, another light element, should also be affected by the polluting material. In particular, and as discussed before, Be is not produced inside stars but only destroyed. Thus the polluting material from the first generation of evolved stars, regardless of the exact nature of the polluting star, is completely devoid of Be. A star formed with only pristine material should have the original Be abundance of that material. \emph{Stars with different amounts of polluted material should have diluted the surface Be abundance to different levels.} Stars with bigger fraction of polluted material should have stronger correlations and anti-correlations between the light elements and should be strongly depleted in Be.

Lithium abundances should in principle behave in a similar way. However, AGB stars can produce Li via the Cameron-Fowler mechanism \citep{1971ApJ...164..111C}. Even though some extreme fine tuning might be needed to explain the full range
of oxygen abundances in stars with seemly normal Li, this option can not be completely excluded as of yet \citep[see e.g.][]{2010A&A...524L...2S}. In this situation, Be offers an important complementary test. As Be is never produced in stars, it is an unique tracer of the amount of pollution suffered by the globular cluster stars.

In this science case for CUBES, the objective is to detect differences in the Be abundance among different turn-off stars of the same cluster. These different stars should have been polluted to different levels. If different Be abundances are detected, then one can use them to quantify the fraction of polluting material. On the other hand, if stars with different levels of pollution are found to have the same Be abundance, this would definitely show that all the simple dilution-pollution scenarios currently proposed to explain the chemical properties of globular clusters are too simplistic and not valid. It would show that the true formation scenario was more complex than that.

Measurements of Be abundances have been attempted in two globular clusters so far: i) In NGC6397, Be was detected in two turn-off stars by \citet{2004A&A...426..651P}, and an upper limit was derived for the extremely Li-rich main-sequence star NGC6397 1657 \citep{2014A&A...563A...3P} and ii) two turn-off stars in NGC6752 were analyzed by \citet{2007A&A...464..601P}, resulting in one detection and one upper limit. The results seem to indicate that the stars have the same Be abundance, even though they have different abundances of oxygen (and thus were polluted to different levels). Nevertheless, the results are uncertain due to the low S/N of the spectra (S/N $\sim$ 10--15), even though they were obtained with long total exposure time ($\sim$ 900 min). CUBES will allow better spectra to be acquired with shorter exposure times, increasing also the number of globular clusters were Be abundances can be investigated. 

To test this science case, synthetic spectra of three different stars were computed. One star represents turn-off stars in the cluster NGC6752, the other stars represent the metal-poor and metal-rich turn-off members of $\Omega$ Cen. For each star, two different Be abundances were adopted. The higher abundance in NGC6752 is the one derived by \citet{2007A&A...464..601P}, log (Be/H) = $-$12.05. The smaller abundance was decreased by 0.60 dex. For the stars in $\Omega$ Cen, the Be abundances were scaled according to the metallicity -- log(Be/H) = $-$12.6 and $-$13.2 for turn-off star 1 and log(Be/H) = $-$11.7 and $-$12.3 for turn-off star 2:

\begin{enumerate}
\item Turn-off star in NGC 6752: $T_{\rm eff}$ = 6400 K; log $g$ = 4.30; [Fe/H] = $-$1.50; $\xi$ = 1.20 km s$^{-1}$
\item Turn-off star 1 in $\Omega$ Cen: $T_{\rm eff}$ = 6700 K; log $g$ = 4.20; [Fe/H] = $-$2.00; $\xi$ = 1.20 km s$^{-1}$
\item Turn-off star 2 in $\Omega$ Cen: $T_{\rm eff}$ = 6550 K; log $g$ = 4.20; [Fe/H] = $-$1.10; $\xi$ = 1.20 km s$^{-1}$
\end{enumerate}

The same combinations of resolving power and S/N used before were adopted. The results are shown in Figs. \ref{fig:n6752}, \ref{fig:cen1}, and \ref{fig:cen2}. It can be seen that basically in all cases a difference in Be abundances can be detected, the only exception being the metal-poor turn-off of $\Omega$ Cen with low-S/N. Therefore, with CUBES it will be possible to use Be as a tracer of the dilution-pollution processes in globular clusters. More sound quantitative predictions have to wait until the CUBES properties are better defined. Then, it will be possible to define the list of clusters were the measurements will be possible and better determine the limit of detection of the Be abundances in these stars.  

\section{Summary}

Abundances of the fragile element Be can be used in different areas of astrophysics, including studies of the Galactic chemical evolution, of stellar evolution, and of the formation of globular clusters. Only the Be II resonance lines at $\lambda$ 3130.422 and 3131.067 \AA\ are used for abundance studies. The lines are in the near UV spectral region, a region strongly affected by atmospheric absorption. It is very time demanding to obtain the high-resolution, high-S/N spectra needed to study Be. 

Because of that, Be abundances have not been used to its full potential yet. To increase the number of stars where Be abundances can be determined, and to decrease the time cost of obtaining high-quality spectra, new UV sensitive instruments are needed. CUBES is one such instrument. It has an expected sensitivity gain of about three magnitudes at 3200 \AA\ over UVES.

In the near future, CUBES will likely be the only instrument offering the opportunity to measure Be abundances in samples of extremely metal-poor stars and in turn-off stars of globular clusters. Here, preliminary simulations of CUBES-like spectra for these types of stars were presented. 

For the first case, the simulations indicate that with CUBES spectra it will be possible to investigate the suggested flattening of the relation between Be and Fe at low metallicities ([Fe/H] $<$ $-$3.00). This flattening might have implications, for example, to primordial nucleosynthesis models.

With CUBES, it will also be possible to investigate Be abundances in turn-off stars of globular clusters. Beryllium can be used as a tracer of the fraction of polluting material accreted by the second generation stars. If stars with different levels of pollution are found to have the same Be abundances, this would argue against the currently favored dilution-pollution scenarios invoked to explain the chemical properties of globular clusters.

%
\acknowledgments
I thank the organizers of the ``Challenges in UV Astronomy'' workshop for the invitation to give the review talk on beryllium abundances. I acknowledge financial support by the National Science Center of Poland through grant 2012/07/B/ST9/04428. I am pleased to thank L. Pasquini and H. Kuntschner for sharing their enthusiasm for the CUBES project.


%
 \bibliographystyle{plain}  
 

%

\end{document}